\PassOptionsToPackage{usenames,dvipsnames}{xcolor}

\documentclass[draft]{agujournal2019}
\usepackage{url} 
\usepackage{lineno}
\usepackage[utf8]{inputenc}
\usepackage{soul}
\usepackage{layouts}
\draftfalse

\journalname{Geophysical Research Letters}

\begin{document}


\title{Spatial and temporal development of incipient dunes}

\authors{C. Gadal\affil{1}, C. Narteau\affil{1}, R.C. Ewing\affil{2}, A. Gunn\affil{3}, D. Jerolmack\affil{3,4}, B. Andreotti\affil{5}, P. Claudin\affil{6}}

\affiliation{1}{Institut de physique du globe de Paris, Universit\'e de Paris, CNRS, Paris, France}
\affiliation{2}{Department of Geology and Geophysics, Texas A\&M University, College Station, USA}
\affiliation{3}{Department of Earth and Environmental Science, University of Pennsylvania, Philadelphia, USA}
\affiliation{4}{Department of Mechanical Engineering and Applied Mechanics, University of Pennsylvania, Philadelphia, USA}
\affiliation{5}{Laboratoire de Physique, ENS - PSL Research University, Universit\'e de Paris, CNRS, Sorbonne Universit\'e, Paris, France}
\affiliation{6}{Physique et M\'ecanique des Milieux H\'et\'erog\`enes, ESPCI Paris - PSL Research University, Universit\'e de Paris, CNRS, Sorbonne Universit\'e, Paris, France}

\correspondingauthor{Cyril Gadal}{gadal@ipgp.fr}

{\vspace{1cm}\color{blue}\small  An edited version of this paper was published by AGU. Copyright 2020 American
Geophysical Union: Gadal, C., Narteau, C., Ewing, R. C., Gunn, A., Jerolmack, D., Andreotti, B., \& Claudin, P. (2020). Spatial and temporal development of incipient dunes. Geophysical Research Letters, 47, e2020GL088919. https://doi.org/10.1029/2020GL088919}.



\begin{keypoints}

\item Length and time scales of dune formation set the pace of aeolian landscape dynamics
\item We provide significant statistics on the temporal and spatial growth of incipient sand dunes extracted from extensive field observations
\item Data provide a field validation of dune instability theory, and introduce the distance of dune growth as an important length scale

\end{keypoints}

%
%

%
%


\begin{abstract}
In zones of loose sand, wind-blown sand dunes emerge due the linear instability of a flat
sedimentary bed. This instability has been studied in experiments and numerical models but rarely in the
field, due to the large time and length scales involved. We examine dune formation at the upwind margin
of the White Sands Dune Field in New Mexico (USA), using 4 years of lidar topographic data to follow
the spatial and temporal development of incipient dunes. Data quantify dune wavelength, growth rate, and
propagation velocity and also the characteristic length scale associated with the growth process. We show
that all these measurements are in quantitative agreement with predictions from linear stability analysis.
This validation makes it possible to use the theory to reliably interpret dune-pattern characteristics and
provide quantitative constraints on associated wind regimes and sediment properties, where direct local
measurements are not available or feasible.
\end{abstract}

\section*{Plain Language Summary}
%
Dunes are the solar system's ubiquitous landform, arising wherever
wind blows over a loose sand bed. An aerodynamic theory for dune formation, which connects grain-scale
movement to emergent dune pattern, has been developed for idealized scenarios. Yet this model has
never been directly tested in nature, because of the complexities in observing dune formation at the initial
stage. Here we report extensive topographic observations of the initiation, growth, and migration of
real-world sand dunes. Moreover, we find a surprisingly precise agreement with the idealized aerodynamic
theory. This robust confirmation of the theory for dune formation means that we may estimate wind
conditions in remote places, including other planets, with confidence.
%

\section{Introduction}
\label{intro}

The development of sand dunes, from incipient to mature bedforms, and their evolution, through interaction and coarsening processes, involve characteristic time and length scales that relate to elementary mechanisms of hydrodynamics and sediment transport \cite{Wigg13, Cour15}. Over loose granular beds, bedform emergence is driven by a hydrodynamic instability induced by the interaction between the sand bed, flow, and sediment transport \cite{Char13}. On the upstream side of a bump, erosion takes place as the flow accelerates. Reciprocally, the flow slows down on the downstream side where deposition occurs. However, the transition between erosion and deposition zones, associated with the location of the maximum of the sediment flux, does not necessarily take place at the crest of the bump. Spontaneous growth of such a bump--that is, instability--can therefore occur if its crest is located in the deposition zone \cite{Kenn63}. The streamwise offset between topography and sediment flux has two contributions \cite{Andr02a, Kroy02a, Four10, Clau13}. First, a hydrodynamic destabilization originates from the coupling between flow inertia and dissipation, which results in a maxi-
mum basal fluid shear stress located upstream of the crest \cite{Syke80, Hunt88, Kroy02b}. Second, the sand flux needs a characteristic length, called the saturation length, to adapt to a spatial change in shear stress \cite{Saue01, Andr10, Dura11, Paeh13}. This results in a stabilizing downstream lag of the maximum sand flux with respect to the maximum of the
shear stress. These balancing processes give rise to the development and propagation of sand waves at a specific wavelength and propagation speed, associated with the most unstable mode of the pattern, with crests perpendicular to the dominant wind direction.

The early stage of growth and development of sedimentary ripples and dunes has been theoretically studied with linear stability analyses of coupled transport and hydrodynamic equations \cite{Kenn63, Rich80, Andr02a, Lagr03, Colo04, Clau06, Four10, Deva10, Andr12, Dura19, Gadal19}. These analyses predict the incipient pattern wavelength, propagation velocity, and growth rate as functions of model parameters, which encode the wind and grain characteristics. For the aeolian case in particular, the dune wavelength has been shown to be proportional to the saturation length. However, measuring the bed elevation together with sediment and fluid transport is difficult, thereby making the direct comparison between theory and field or experimental data rather challenging.

The aerodynamic and sediment transport responses have been investigated independently of each other, and separate measurements of the saturation length and the upwind shift of the shear stress have been carried out, in the field and in wind tunnel experiments \cite{Andr10, Clau13, Selm18}. In contrast, few field studies addressing the early stage of aeolian dune growth are available in the literature \cite{Cooper1958, Fryberger1979, Kocurek1992, Elbe05, Ping14, Badd18}. First, in situ monitoring of the evolution of small amplitude bedforms is difficult due to the involved length and time scales (tens to hundreds of meters, days to months). Second, inherent wind variability--even in overall unidirectional dune fields--makes application of the theory challenging. Emergence of subaqueous sand ripples has also been experimentally investigated \cite{Cole96, Baas99, Lang07, Four10}, and more generally the quantification of sedimentary bedforms in different environmental--including extraterrestrial--conditions in relation to hydrodynamics and sediment transport remains an active current subject of research \cite{Lapo16, Jia17, Lapo18, Dura19, Gadal19}.

In this paper, we study the upwind margin of the White Sands Dune Field, where the dune instability leads
to spatially amplifying sand waves developing downstream \cite{Ewin10, Phil19}. We follow the spatiotemporal evolution of incipient dunes and extract their wavelength and propagation velocity, as well as their temporal and spatial growth rates. We then show that these four quantities all quantitatively compare to the predictions of spatial linear stability analysis.

\begin{figure}
  \centering
   \includegraphics[width = \textwidth]{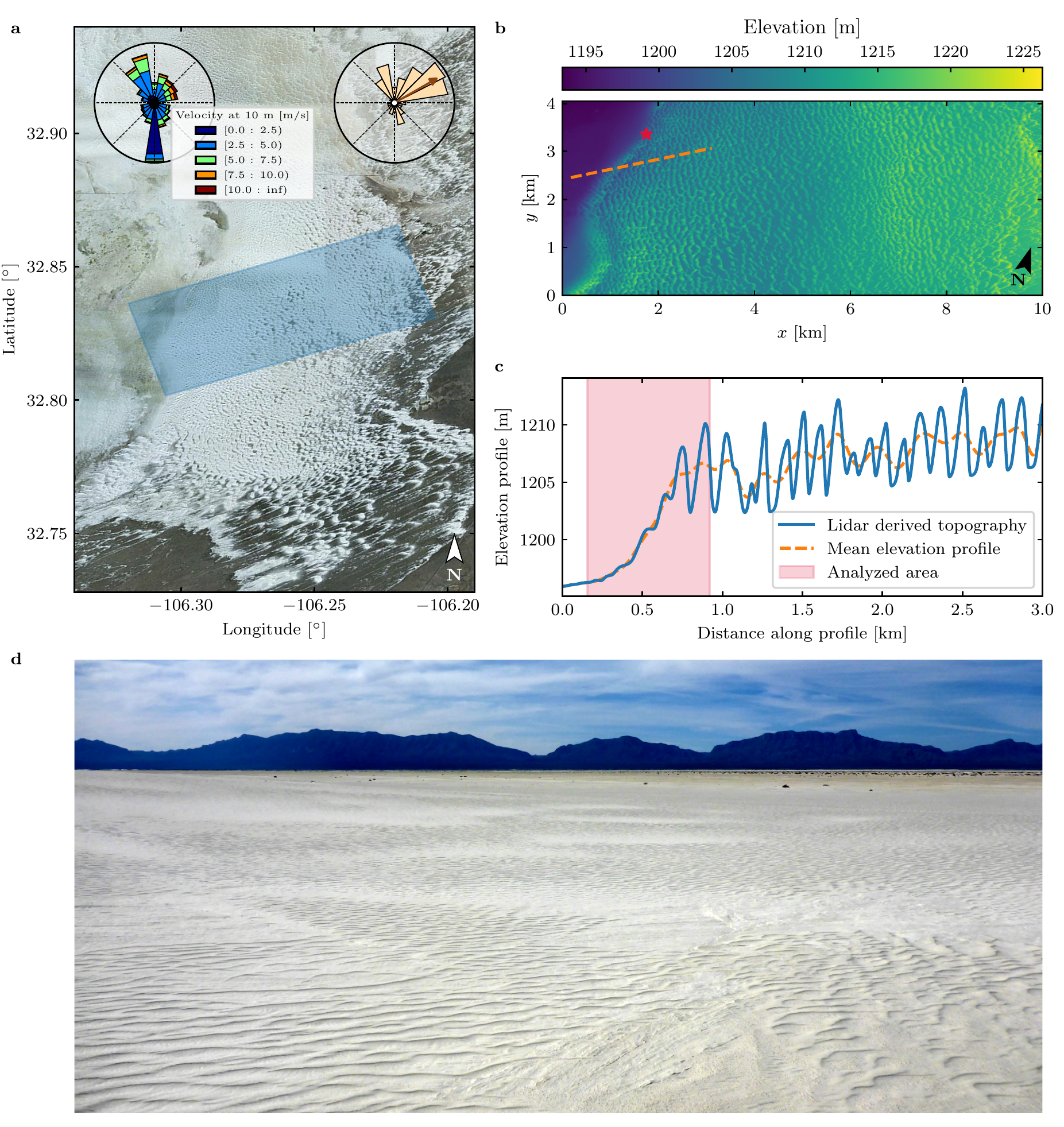}
  \caption{The White Sands National Park dune field. (a) Satellite image of the White Sands Dune Field (Google$^{\rm TM}$, Landsat-Copernicus). The left rose shows the wind data from 2007 to 2017 (direction toward which the wind is blowing). The right rose shows the corresponding distribution of sand flux orientations and the resultant flux direction (brown arrow). Both agree well with that reported by \citeA{Ewing2015, Pede15}. The blue area corresponds to the location of the digital elevation measurements. (b) Digital elevation data taken in June 2007. The dashed orange line is the location of the transect shown in (c), taken along the direction of the resultant flux. The red star is the location of the photo shown in (d), which is a view to the southwest of the dune field upwind margin.}
  \label{Figure_terrain}
\end{figure}

\section{White Sands Dune Field}
\label{WSDF}

White Sands Dune Field is located in southern New Mexico, USA. The sand covers an area of about $400~\textrm{km}^{2}$, resulting in the largest gypsum dune field on Earth. Dominant winds are mainly toward the northeast and shape the sedimentary bed into transverse and barchan dunes, progressively turning into parabolic dunes as the vegetation cover increases (Figures~\ref{Figure_terrain}a and~\ref{Figure_terrain}b) \cite{McKee1966, Jero12, Bait14}. Dunes emerge on the upwind margin (Figure~\ref{Figure_terrain}c). There, the sediment is made of coarse, elongate, and angular grains (see Figure~S2 in the supporting information), with a measured diameter $d = 670 \pm 120~\mu\textrm{m}$ and a bulk density $\rho_{\rm p} = 2300 \pm 100~\rm{kg~m^{-3}}$. The saturation length, relevant in the process of dune emergence (see section~\ref{LSA}), directly depends on these grain properties \cite{Andr10}:
\begin{linenomath*}
\begin{equation}
  L_{\rm sat} \simeq  2.2 \, \frac{\rho_{\rm p}}{\rho_{\rm f}} d = 2.8 \pm 0.5 ~\textrm{m},
  \label{Lsat}
\end{equation}
\end{linenomath*}
where $\rho_{\rm f} \simeq 1.2~\rm{kg~m^{-3}}$ is the air mass density in ambient conditions. As one moves further into the dune field, the grain diameter and angularity both decrease, due to abrasion and aeolian sorting \cite{Jero11, Phil19}. Because we restrict our analysis of the dune development to the first kilometer along the margin, the grain characteristics are assumed to be homogeneous. The grain roughness leads to a measured avalanche slope $\mu = 0.8 \pm 0.05$ (see supporting information section S1).

The sand flux is calculated from the hourly wind data of the weather station at Holloman Air Base (KHMN, $32^{\circ}51'$N, $106^{\circ}06'$W), using the method described in \citeA{Cour14} (see also supporting information section S2). The wind is characterized by its shear velocity $u_*$, representative of the logarithmic profile inside the turbulent boundary layer. Its threshold value $u_{\rm th}$ below which saltation cannot sustain steady transport is estimated with $u_{\rm th} = a \sqrt{\left(\rho_{\rm p}/\rho_{\rm f} \right) gd} = 0.35~\textrm{m}~\textrm{s}^{-1}$, and $a \simeq 0.1$ \cite{Iver99}. The corresponding typical velocity ratio $u_{*}/u_{\rm th}$ is then about $1.26 \pm 0.05$ (all these values are gathered in Table~S1). As shown in Figure~\ref{Figure_terrain}a and documented by \citeA{Pede15}, the wind regime is multimodal. Southwesterly winds dominate the transport as noted by the nearly unimodal sand flux distribution toward the northeast. The other modes from the north and southeast moderately impact the dune shape and migration\cite{Swan16}.

\begin{figure}
  \centering
   \includegraphics[width = \textwidth]{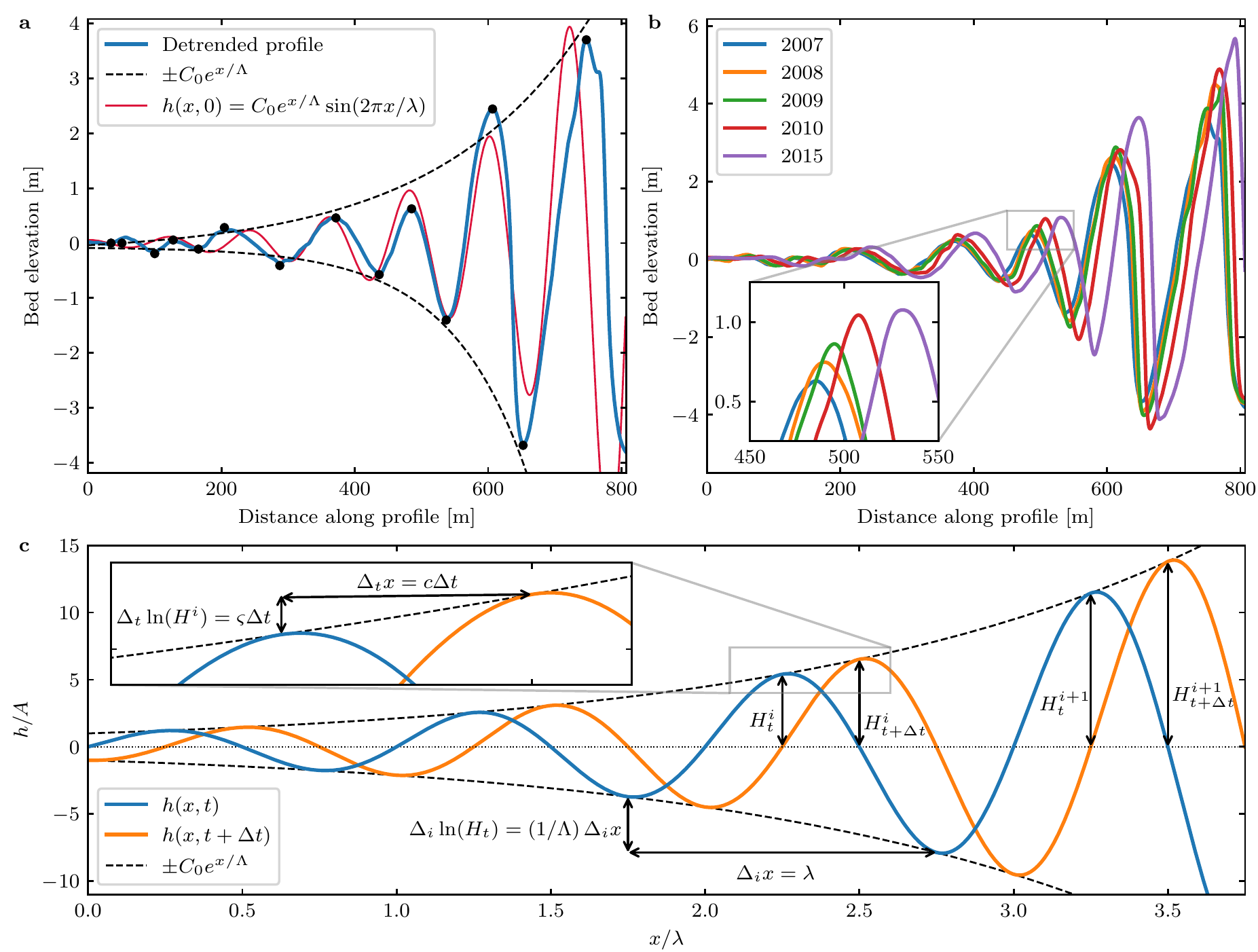}
  \caption{The spatial exponential dune growth. (a) Detrended profile corresponding to that of Figures~\ref{Figure_terrain}b and \ref{Figure_terrain}c. The black dashed lines are exponential fits to the dune crests (black dots), giving $\Lambda = 225 $~m (top) and $\Lambda = 135 $~m (bottom). The theoretical red profile parameters are $C_{0} = 0.06$~m, $\lambda = 120 $~m, and $\Lambda = 170 $~m. (b) Temporal evolution of the detrended elevation profile, with a close-up on one crest. (c) Schematics of the spatio-temporal dune development. The theoretical profile $h$ is defined in equation.~(\ref{spatial_profile}).}
  \label{Figure_analyse}
\end{figure}

Elevation data of the blue area in Figure~\ref{Figure_terrain}a have been obtained using lidar-derived topography at five different times (June 2007, June 2008, January 2009, June 2010 and August 2015). Along the upwind margin, we extracted 75 dune profiles from the surface elevation data, spaced $50$-m apart and aligned with the direction of the resulting sand flux (Figure~\ref{Figure_terrain}b). The average topography is removed using a Butterworth high-pass filter ($\textrm{order = 5}$, $\textrm{cutoff frequency} = 0.005~\textrm{m}^{-1}$). We limit our analysis to the first dunes of the filtered profile to ensure that we stay as much as possible in the early stage of dune growth (red area in Figure~\ref{Figure_terrain}c). As shown by figure~\ref{Figure_terrain}d, these incipient dunes have very low aspect ratios and do not exhibit any slipfaces \cite{Phil19}. The detrended bed elevation exhibits a spatially amplified oscillating behaviour (Figure~\ref{Figure_analyse}a and~\ref{Figure_analyse}b). We now interpret these profiles using the theoretical framework provided by the linear stability analysis developed in the next section.

\section{Dune linear stability analysis}
\label{LSA}

Here, we consider a unidirectional wind blowing at a constant shear velocity $u_{*}$, over a flat sedimentary bed. Above the transport threshold velocity $u_{\rm th}$, the saturated sediment flux induced by this flow follows a quadratic law:
\begin{linenomath*}
\begin{equation}
  q_{\rm sat} = \Omega \left(u_{\rm *}^{2} - u_{\rm th}^{2}\right),
\end{equation}
\end{linenomath*}
where $\Omega$ is a dimensional constant that depends on fluid and grain properties \cite{Unga87, Iver99, Crey09, Dura11}. In natural conditions, however, a sedimentary bed is never perfectly flat nor infinite, and these irregularities can be seen as the sum of different perturbations. The purpose of the linear stability analysis is precisely to study the temporal or spatial evolution of the bed in response to a perturbation of a given time or length scale. The emerging dune pattern is then expected to be dominated by the most unstable scale, associated with a sinusoidal mode of wavelength $\lambda$ and propagation velocity $c$, and whose amplitude grows in time with a rate $\sigma$ or in space over a length $\Lambda$.

Above such a sinusoidal bed, wind and sand flux are also modulated. As described in the introduction, the basal shear stress is not in phase with the topography; this is quantified with two dimensionless coefficients, $\mathcal{A}$ and $\mathcal{B}$, which represent the in-phase and in-quadrature components, respectively. They are weak (logarithmic) functions of $\lambda$ \cite{Four10, Char13} but can be, in the first approximation for our purpose, considered as constant parameters of the model. The corresponding upwind shift of the wind with respect to the bed elevation is $\sim \lambda\mathcal{B}/\left(2\pi\mathcal{A}\right)$. Similarly, the actual sediment flux $q$ is not saturated but delayed with respect to the basal shear stress, a process quantified by the saturation length $L_{\rm sat}$. These are the main physical mechanisms of the dune formation model from which the stability analysis is derived (see supporting information section S3 for the proper technical derivation and related theoretical figures).

\label{temporal_LSA}
Consider first a large flat extent of sand. Under the action of the wind, dunes emerge everywhere simultaneously: there is no spatial development of the pattern. A spatial sinusoidal perturbation characterized by a given wavelength $\lambda$ and an initial amplitude $C_0$ can grow or decay in time, at a rate $\sigma$, in response to the wind shift and the flux lag. The perturbation also propagates at a velocity $c$. The bed elevation along the direction $x$ of the wind can be written as:
\begin{linenomath*}
\begin{equation}
  h(x,t) = C_{0} \, e^{\sigma t} \cos\left[\frac{2\pi}{\lambda}\left(x-c t\right)\right].
  \label{temporal_profile}
\end{equation}
\end{linenomath*}
Both temporal growth rate and propagation speed can be computed as functions of $\lambda$ from the analysis of the equations coupling the flow, sediment transport and bed evolution, constituting the dispersion relation of sand waves (supporting information section S3). Positive values of the growth rate are associated with unstable perturbations, and these are typically for large values of $\lambda$. Conversely, small wavelengths are stable, with $\sigma<0$.

\label{spatial_LSA}
Consider now a sediment bed bounded upwind such that disturbances cannot grow at a specific position in space, noted here $x=0$. Dunes emerge from the selective amplification of disturbances propagating downwind from the field entrance. This results in a spatial development of the pattern. There is no temporal growth: at a given location, the amplitude of the bed oscillation is the same at any time. The form of a sinusoidal mode of initial amplitude $C_0$ can be written as:
\begin{linenomath*}
\begin{equation}
  h(x,t) = C_{0} \, e^{x/\Lambda} \cos\left[\frac{2\pi}{\lambda}\left(x-c t\right)\right],
  \label{spatial_profile}
\end{equation}
\end{linenomath*}
where $\Lambda^{-1}$ is the spatial growth rate of the dunes.

Neutral modes are the same in both spatial and temporal analysis. Their wavelength $\lambda_c$ is characterised by vanishing growth rates $\sigma(\lambda_c)=0$ and $\Lambda^{-1}(\lambda_c)=0$, such that,
\begin{linenomath*}
\begin{equation}
  \lambda_{c} = \displaystyle\frac{2\pi\mathcal{A}}{\mathcal{B} - \displaystyle\frac{1}{\mu}\left(\displaystyle\frac{u_{\rm th}}{u_{*}}\right)^{2}} \, L_{\rm sat},
\label{lambdac}
\end{equation}
\end{linenomath*}
and separates growing ($\lambda > \lambda_c$) from decaying ($\lambda < \lambda_c$) perturbations. It can thus be interpreted as a minimal dune size.

Performing the temporal linear stability analysis (denoted by subscript T), in the limit $L_{\rm sat}/\lambda_c \ll 1$, the characteristics of the fastest growing perturbation read:
\begin{linenomath*}
\begin{eqnarray}
	\lambda_{\rm T} & \sim & \frac{3}{2} \lambda_{c},
    \label{def_lambda} \\
	\sigma_{\rm T} & \sim & \frac{Q}{L_{\rm sat}^2} \, \frac{\mathcal{A}}{2} \, \left(\frac{2 \pi L_{\rm sat}}{\lambda_{\rm T}}\right)^{3},
    \label{def_sigma} \\
	c_{\rm T} & \sim & \frac{Q}{L_{\rm sat}} \, \mathcal{A} \, \frac{2 \pi L_{\rm sat}}{\lambda_{\rm T}},
  \label{def_c}
\end{eqnarray}
\end{linenomath*}
where $Q = \Omega u_{\rm *}^2$ gives the characteristic scale of the sediment flux associated with the wind regime \cite{Four10, Gadal19}.

Conversely, spatial growth rate reaches a maximum $1/\Lambda_{S}$ at some specific value of the wavelength, noted $\lambda_{\rm S}$, corresponding to the propagation velocity $c_{\rm S}$. Unfortunately, no simple analytical and accurate formulae like (\ref{def_lambda}-\ref{def_c}) can be derived for these quantities (see supporting information section S3). Temporal and spatial analyses are consistent, and we typically find $\lambda_{\rm S} \simeq 1.3 \lambda_{\rm T}$ and $c_{\rm S} \simeq 0.77 c_{\rm T}$. The numerical factors in these relations do not vary by more than a few percent upon changing the parameters $\mathcal{A}$ and $\mathcal{B}$ within a reasonable range of values. In the spatial development of the instability, an individual bump grows in height while propagating downwind at a constant velocity. Its amplitude therefore varies exponentially with respect to time, and one can define a temporal growth rate $\varsigma_{\rm S} \equiv c_{\rm S}/\Lambda_{\rm S}$ (inset of Figure~\ref{Figure_analyse}c) in the frame of reference of the bump i.e. a Lagrangian growth rate. The theoretical analysis provides the approximate relation $\varsigma_{\rm S} \simeq 0.43\sigma_{\rm T}$. Both $\varsigma_{\rm S}$ and $\Lambda_{S}$ can be measured from the field data, providing equivalent information.

\section{Field data analysis}
After removing the average topography (Figure~\ref{Figure_terrain}c), the detrended bed elevation profile exhibits an exponentially amplifying sinusoidal shape, as predicted by the spatial linear theory for dune emergence (Figure~\ref{Figure_analyse}a). Using these profiles and their temporal evolution (Figure~\ref{Figure_analyse}b), we have extracted the three independent characteristics of the pattern ($\lambda$, $c$ and $\Lambda$ or $\varsigma$) using two different methods. For each profile, we either look at each peak separately (peak-to-peak method), or extract quantities averaged over the whole profile (global approach).

The wavelength $\lambda$ is computed by autocorrelation of the bed elevation profile (global method), and from the spacing between two adjacent peaks (peak-to-peak method). The fit of an exponential to the peaks of each profile gives the spatial growth length $\Lambda$ (global method, see dashed line in Figure~\ref{Figure_analyse}a). The spatial growth length is alternatively computed from the difference in height between two adjacent peaks (peak-to-peak method, see Figure~\ref{Figure_analyse}c).

The Lagrangian growth rate $\varsigma$ and propagation velocity $c$ are obtained by fitting exponential and linear functions to the temporal variation of the dune height and position, respectively (see equations 43 and 44 of the supporting information section S3). The peak-to-peak method looks at the height and position of each peak separately (see inset of Figure~\ref{Figure_analyse}c). For a global measurement, the average propagation speed can be determined from the cross-correlation curve between the same profiles at different times, and the average Lagrangian growth rate from the temporal evolution of the bed elevation standard deviation.

Importantly, to extract the average values of $\varsigma$ and $c$, we take into account the temporal variations of the characteristic sand flux $Q$. Indeed, both quantities vary in time proportionally to $Q$.  We also remove periods of time when the wind is below the transport threshold (see supporting information section S4).


\section{Time and length scales of the incipient dunes}
\label{distrib}

\begin{figure}
  \centering
   \includegraphics[width = \textwidth]{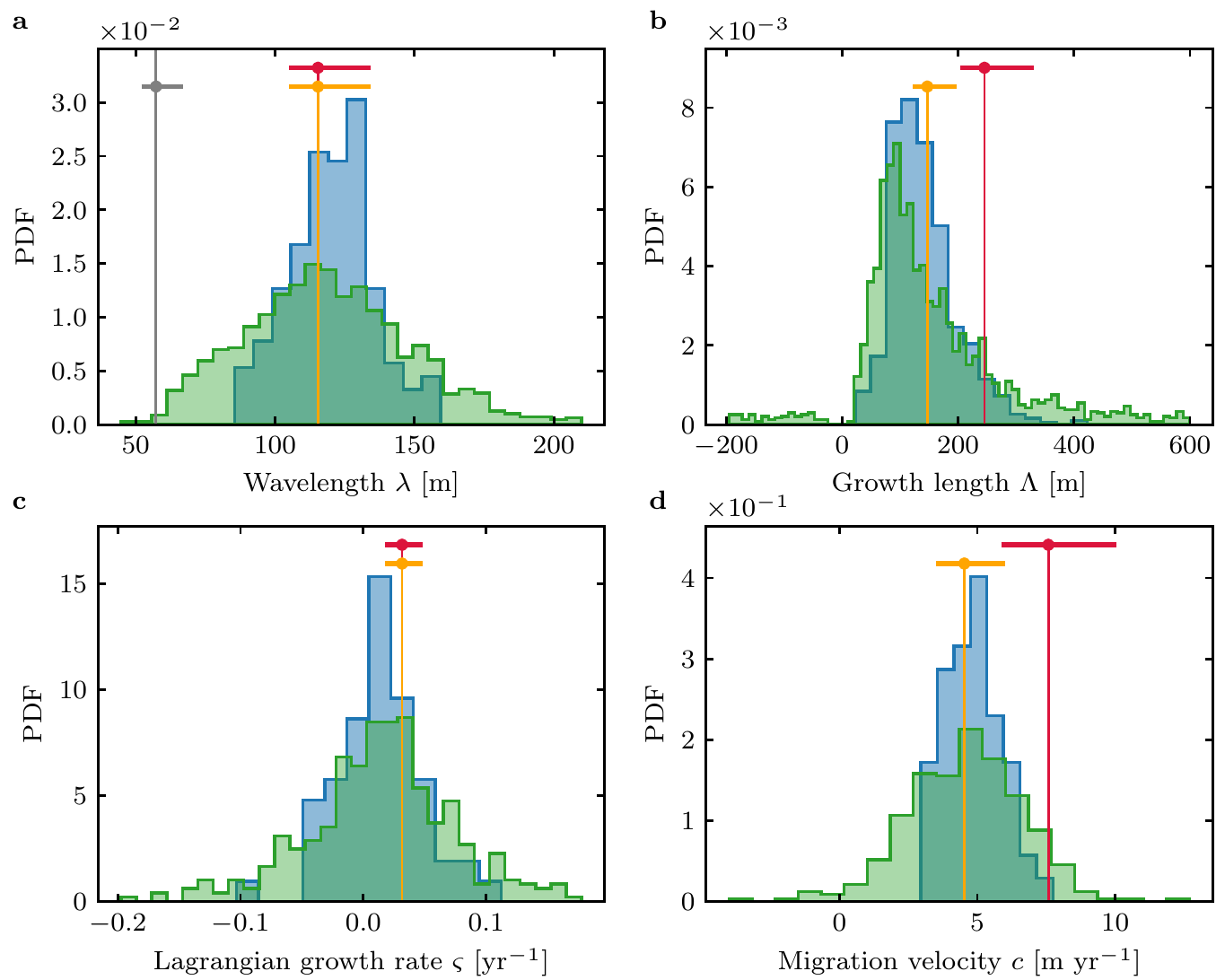}
  \caption{Distributions of incipient dune time and length scales. Blue and green distributions shows the results of the global and peak-to-peak methods, respectively. Errorbars gives the range of values obtained from the spatial linear stability analysis with $\mathcal{A} = 3.6 \pm 0.6$ and $\mathcal{B} = 1.9 \pm 0.3$, and dots shows the average. Raw predictions based on a unidirectional wind are in red, and predictions taking into account the correction due to reversing winds are in orange (see Table~S2). The cut-off wavelength $\lambda_{c}$ is in grey.}
  \label{Figure_distributions}
\end{figure}

The output of the analysis of the $75$ transects is shown in Figure~\ref{Figure_distributions}. Both peak-to-peak and global methods exhibit similar distributions for the wavelength $\lambda$, the propagation velocity $c$, the Lagrangian growth rate $\varsigma$ and the growth length $\Lambda$, with clear dominant (most probable) values. The incipient dune wavelength and growth length are both on the order of a hundred meters; their propagation velocity is around $5$~m~yr$^{-1}$ and their growth rate is about $0.015$~yr$^{-1}$. These values, as well as the typical dispersion around them (i.e. the width of these distributions), are more precisely reported in Table~S2. For $\lambda$ and $c$, our results are consistent with the measurements of \citeA{Phil19}, made on a single elevation profile. Their dispersion is on the order of $20$\%, because these quantities can be measured with a good accuracy, especially with the global method using correlation. As the measurement of $\Lambda$ and $\varsigma$ is more delicate, the corresponding distributions are more dispersed. The peak-to-peak method is actually sensitive to the behaviour of individual peaks, that can respond to various types of local disturbances. For example, they may induce irregularities in the spacing between the peaks, or asymmetry between positive and negative detrended topography (Figure~\ref{Figure_analyse}b). As a result, a few negative values of the velocity, growth rate and characteristic growth length are reported. Nevertheless, these data provide reliable and meaningful statistics to test the theory, which must be able to account for those four quantities concomitantly. The free parameters of the linear analysis are the hydrodynamic coefficients $\mathcal{A}$ and $\mathcal{B}$, as the others are set independently with the wind and sediment properties (see Table S1).

The incipient dune wavelength peaks around $120$~m, and is therefore significantly larger than the usually reported value ($\simeq 20$~m, for sand particles of size $180~\mu$m) \cite{Elbe05}. How can the theory reproduce such a large value? First, as the wavelength is proportional to the saturation length, and thus to the grain size (equations~\ref{Lsat},~\ref{lambdac} and~\ref{def_lambda}), the presence here of much coarser grains provides a factor $670/180 \simeq 3.7$ corresponding to the ratio of grain diameter. Second, the most unstable wavelength is predicted to increase when sediment transport occurs close to the threshold, in relation to the denominator of equation~\ref{lambdac} \cite{Andr10, Char13, Gadal19}. Here, with $u_{*}/u_{\rm th} \simeq 1.3$, we can expect an additional factor of $2$ with respect to a situation far above threshold. The simultaneous fit of the four quantities predicted by the spatial linear stability analysis to the data allows us to reproduce quantitatively the dominant wavelength and growth rate. The predicted values fall in the peak of the distributions (red lines in Figures~\ref{Figure_distributions}a and \ref{Figure_distributions}c). This adjustment however overestimates the growth length and the propagation velocity, whose predictions exceed the dominant values by an amount comparable to the peaks' width (red lines in Figures~\ref{Figure_distributions}b and~\ref{Figure_distributions}d).

This discrepancy can be understood by questioning our approximation of a unidirectional wind. A finer analysis of the flux rose shows in fact secondary winds, with non-zero components perpendicular to the crest toward the southwest. Reversing winds have cumulative effects on the growth rate and the selected wavelength. They however partially cancel each other out, impeding the propagation, and thus the spatial development of the dune pattern. Such a process is supported by observations of reactivation surfaces formed by reversing winds in the stratigraphy at White Sands \cite{Phil19, Kocu07}. The value of the characteristic sand flux $Q$ (given in Table~S1) is a time average that does not account for changes in wind orientation. The ratio between the scalar and vector averages of $Q$, taking into account the variations in orientation of the sand fluxes over time, is about $0.6$ (see supporting information section S4). Once corrected by this ratio, the predicted migration velocity and growth length come into quantitative agreement with the corresponding dominant values of the distributions (orange lines in Figure~\ref{Figure_distributions}).

\section{Discussion}
\label{discuss}

The exponentially amplified sinusoidal behaviour of the White Sands bed elevation profiles is field evidence for the spatial development of the dune instability. The linear analysis is able to quantitatively reproduce the three characteristics of the emergent pattern: dune wavelength, propagation velocity and growth length (or equivalently, growth rate). To obtain this agreement, the two hydrodynamic coefficients where adjusted, resulting in $\mathcal{A} = 3.6 \pm 0.6$ and $\mathcal{B} = 1.9 \pm 0.3$. All the other parameters of the theory were fixed independently from sediment and wind data, either by direct measurement (grain diameter and density, avalanche slope) or using well calibrated relationships (saturation length, sediment flux). Note that the uncertainty on the determined coefficients is dominated by the those of the fixed parameters, rather than by the dispersion in the measurements of $c$, $\lambda$ and $\Lambda$. Overall, the value of $L_{\rm sat}$, involved in $\lambda$ and $\Lambda$, directly affects the estimate of $\mathcal{B}/\mathcal{A}$, while the value of $Q$, involved in $c$, mostly affects $\mathcal{A}$.

Importantly, the concomitant agreement of $c$, $\lambda$ and $\Lambda$ is a stringent test of the theory. This study is therefore a step forward in the general `dune inverse problem', trying to infer, for example, grain or wind properties from dune characteristics \cite{Fent14a, Fent14b, Ewing2015, Runy17, Fern18}. It is remarkable that the resulting values determined from our field data are very close to those directly measured by \citeA{Clau13} on a single $40$ m long dune ($\mathcal{A} \simeq 3.4$ and $\mathcal{B} \simeq 1.6$), as well as to the predictions of hydrodynamic models \cite{Four10, Char13}. This study thus confirms the reliability of the linear analysis in the interpretation of field, experimental or numerical data relating to the emergence of incipient dunes \cite{Elbe05, Nart09, Four10, Gadal19}.

A limitation of the linear theory is of course the presence of non-linear effects. They occur when the aspect ratio of the sand waves becomes too large or when the dunes interact with each other, so that each bed perturbation cannot be considered as independent of the others. Bumps with aspect ratios above $\simeq 1/13$ are expected to start to develop flow recirculation on their downwind side, usually associated with the formation of an avalanche slip face \cite{Four10}. In the region we have studied, the waves furthest inside the dune field could reach aspect ratios of about $1/10$, but no slip faces were observed. Similar to \citeA{Phil19}, we also recognize the coexistence of multiple wavelengths at the upwind side of the profiles (associated with different celerities and growth rates or lengths), and these are partly the cause of the distribution widths in Figure~\ref{Figure_distributions}. We could not, however, infer from these data signs of interactions, such as collisions, coalescence or ejection \cite{Hersen2005, Katsuki2005, Gao15b, Bacik2020}.

Although studied here at the boundary of a dune field, the spatial development of the dune instability is also present on pre-existing large dunes, providing a similar upwind boundary condition in terms of sand availability. As a matter of fact, the emergence of bed oscillations on the flanks of barchans has been proposed as a key mechanism to understand their stability, as these superimposed waves eventually grow until they can break from the horns, causing large sand losses \cite{Elbe05, Zhan10, Lee2019}. Likewise, in narrow bidirectional wind regimes, the growth of the instability over elongating linear dunes breaks them into trains of barchans \cite{Gao15}. This work therefore provides a reliable base to study the stability of large dunes and thus the formation of large-scale structures inside dune fields \cite{Worm13, Gadal20}.

%
%
%
%
%
%
%
%

\acknowledgments
This collaborative work was initiated at the International Conference on Aeolian Research
(ICAR), 2018, Bordeaux, France. We acknowledge financial support from the Univ-EarthS LabEx program of Sorbonne Paris Cit\'e (grants ANR-10-LABX-0023 and ANR-11-IDEX-0005-02) and the French National Research Agency (grant ANR-17-CE01-0014/SONO). Cl\'ement Narteau acknowledges support from the National Science Center of Poland (grant 2016/23/B/ST10/01700). Philippe Claudin acknowledges support from TOAD (The Origine of Aeolian Dunes) project as an external partner (Natural Environment Research Council, UK and National Science Foundation, USA; NE/R010196NSFGEO-NERC). Ryan C. Ewing acknowledges support from the White Sands National Monument through NPS-GC-CESU Cooperative Agreement \#P12AC51051.
Meteorological data used in this manuscript are hosted by MesoWest under the station code `KHNM' (\url{https://mesowest.utah.edu/cgi-bin/droman/mesomap.cgi?state=NM&rawsflag=3}).
Topographic data can be found in the public repositories Texas Data Repository and OpenTopography (2007-doi:10.18738/T8/WUNF0G, 2008-doi:10.18738/T8/HQVUSX, 2009-doi:10.5069/G9Q23X5P, 2010-doi:10.5069/G97D2S2D, 2015:\url{https://portal.opentopography.org/usgsDataset?dsid=USGS_LPC_NM_WhiteSands_2015_LAS_2017}).
Supplementary figures, tables, and text can be found in the supporting information. We thank the anonymous reviewers for their careful reading of our manuscript and their insightful comments and suggestions.

\bibliography{biblio.bib}

%

\end{document}


\title{Supporting Information for ``Spatial and temporal development of the dune instability"}


\authors{C. Gadal\affil{1}, C. Narteau\affil{1}, R.C. Ewing\affil{2}, A. Gunn\affil{3}, D. Jerolmack\affil{3}, B. Andreotti\affil{4}, P. Claudin\affil{5}}

\affiliation{1}{Institut de physique du globe de Paris, Universit\'e de Paris, CNRS, Paris, France}
\affiliation{2}{Department of Geology and Geophysics, Texas A\&M University, College Station, USA}
\affiliation{3}{Department of Earth and Environmental Science, University of Pennsylvania, Philadelphia, USA}
\affiliation{4}{Laboratoire de Physique, ENS - PSL Research University, Universit\'e de Paris, CNRS, Sorbonne Universit\'e, Paris, France}
\affiliation{5}{Physique et M\'ecanique des Milieux H\'et\'erog\`enes, ESPCI Paris - PSL Research University, Universit\'e de Paris, CNRS, Sorbonne Universit\'e, Paris, France}

\begin{article}

\newpage

\noindent\textbf{Contents of this file}
\begin{itemize}
\item Introduction
\item Data availability statement
\item Text S1 to S5:
\begin{enumerate}
  \item Text S1. Measurement of the avalanche angle
  \item Text S2. Wind data processing
  \item Text S3. Linear stability analysis
  \item Text S4. Temporal fit of the measured quantities
  \item Text S5. Distributions of the hydrodynamic coefficients and uncertainties
\end{enumerate}
\item Figures S1 to S6
\item Tables S1 to S2
\end{itemize}

\newpage

\noindent\textbf{Introduction}

In this supporting information, we provide details on data measurement and processing for the avalanche angle (Text S1) and wind data (Text S2). We also derive the temporal and spatial linear stability analysis (Text S4). The spatial analysis technically differs from the standard temporal analysis and has never been published before in the literature. We also explain how to appropriately scale time when comparing data and theory when sediment transport is intermittent (Text S4), and how the hydrodynamic parameters in the analysis can be estimated from the data (Text S5). This document is prefaced with information on where to find the data, and ends with supporting figures pertaining to Text S1-S5 (Figures S1-S6) and tables summarizing all numerical quantities useful for our analysis (Tables S1-S2).


\vspace*{0.3cm}

\noindent\textbf{Data availability statement}

The wind data is analyzed for the period 2007-2010, based on the data collected by the United States Air Force, and hosted by MesoWest under the station code `KHNM' here: \url{https://mesowest.utah.edu/cgi-bin/droman/mesomap.cgi?state=NM&rawsflag=3}. All of the avalanche slope data is reported in Figure S1. The topographic data information is as follows:
\begin{itemize}
    \item June 2007 (DOI:10.18738/T8/WUNF0G) hosted by Texas Data Repository, collected by Gary Kocurek, and funded by the National Park Service.
    \item June 2008 (DOI:10.18738/T8/HQVUSX) hosted by Texas Data Repository, collected by Gary Kocurek, and funded by the National Park Service.
    \item January 2009 (DOI:10.5069/G9Q23X5P) hosted by OpenTopography, collected by Gary Kocurek, and funded by the National Center for Airborne Laser Mapping.
    \item June 2010 (DOI:10.5069/G97D2S2D) hosted by OpenTopography, collected by Ryan Ewing, and funded by the National Center for Airborne Laser Mapping.
    \item August 2015 is hosted by OpenTopography and collected as part of the United States Geological Survey 3D Elevation Program. Available here: \url{https://portal.opentopography.org/usgsDataset?dsid=USGS_LPC_NM_WhiteSands_2015_LAS_2017}.
\end{itemize}

\vspace*{0.3cm}

\noindent\textbf{Text S1. Measurement of the avalanche slope $\mu$}

\begin{figure}
  \centering
  \noindent\includegraphics[scale = 1, draft = false]{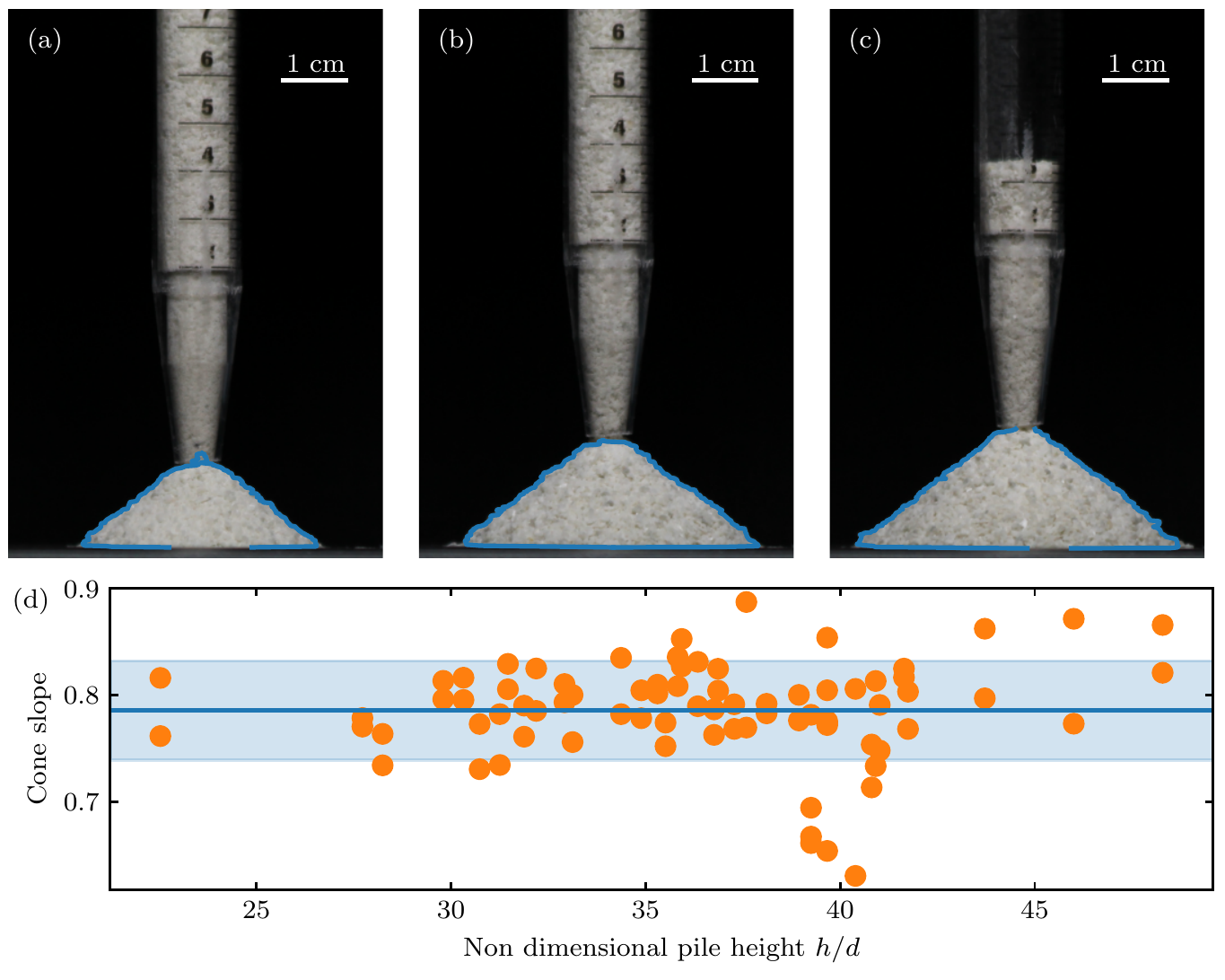}
  \caption{(a-c) Evolution of the sand pile when the tube is pulled up. Contour detection is shown by the blue lines. (d) Measured slope as a function of the non dimensional pile height, where $d = 435$~$\mu$m is the $d_{50}$ of the particular sample we have used. The blue line is the mean value, and the shaded area represents one standard deviation above and below it.}
  \label{Figure_sand_pile}
\end{figure}

\begin{figure}
  \centering
  \noindent\includegraphics[scale = 1, draft = false]{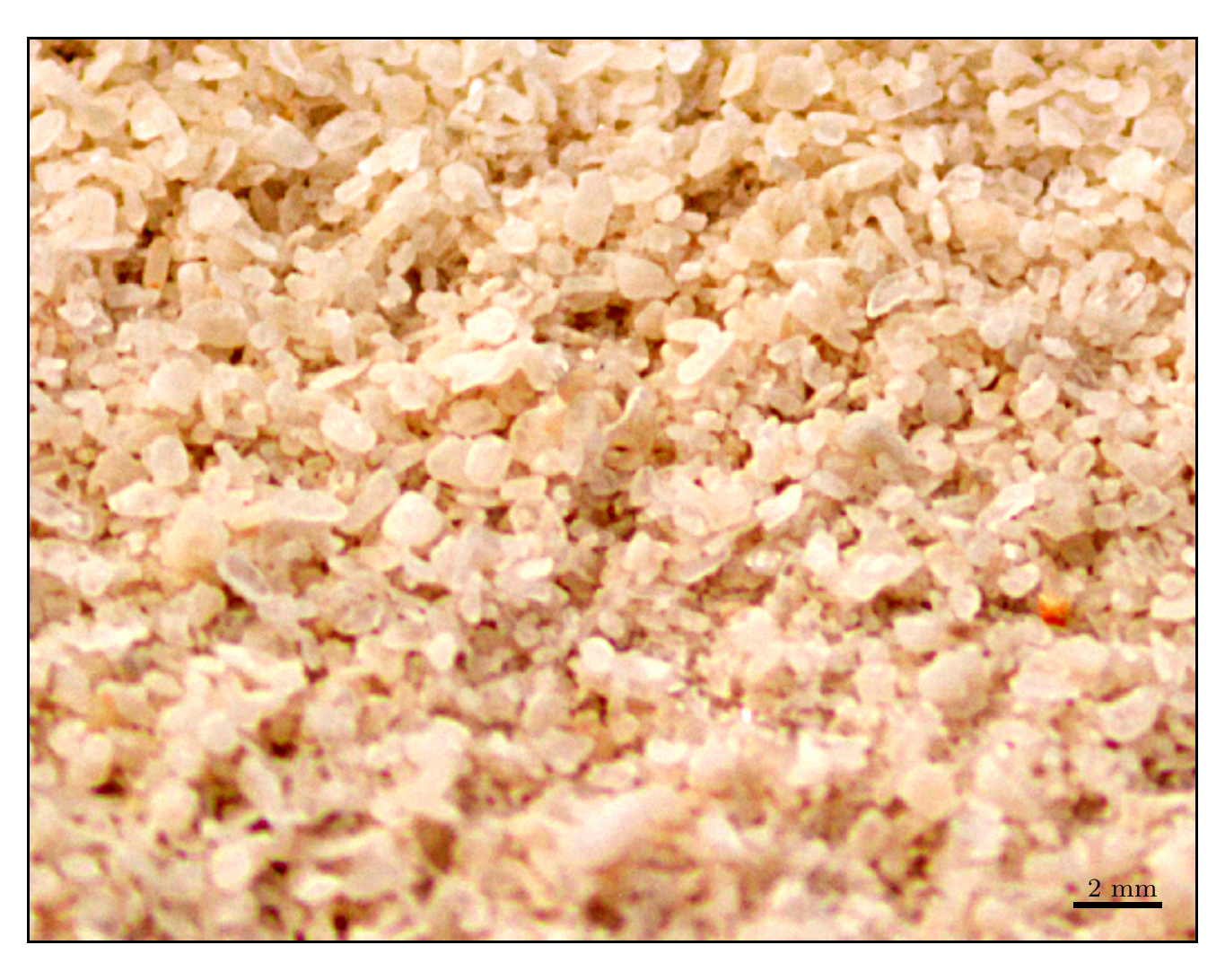}
  \caption{In situ image of representative gypsum grains on the upwind margin; their oblate shape and large size can be seen. The contrast and brightness of grains have been adjusted so their shapes are identifiable.}
  \label{Figure_field}
\end{figure}

The avalanche slope $\mu$ of a granular material is measured from the shape of a conical pile built with these grains. Its value slightly depends on the grains' shape, and, because it enters the parameters of the linear stability analysis, we have measured it for the gypsum particles that compose the dunes at the the upwind margin of the White Sands. The experimental set-up is presented in Figure~\ref{Figure_sand_pile}. We fill a cylindrical tube with the grains and slowly pull it up. The grains flow out of the tube bottom, forming, in a quasi-static way, a conical pile. The slope of the pile evolves through a succession of avalanches. We take pictures after every incremental lift of the tube. We have so obtained data for $39$ cones of different volumes. On these pictures, we detect the edges of the pile by simple image processing, and fit them by straight lines to measure the corresponding slope. The results are shown in Figure~\ref{Figure_sand_pile}. The values are nicely independent of the pile height, as it should for homothetic cones, showing that, finite size effects (the piles are small) can be neglected. These data allow us to obtain a mean value:
\begin{equation}
  \mu = 0.79 \pm 0.05,
\end{equation}
corresponding to an avalanche angle $\simeq 38^\circ$. It is slightly larger than the usual value of $35^\circ$, in agreement with the elongated and angular shape of the grains (see figure~\ref{Figure_field}). The dispersion of the data relates to the metastability of the avalanching process.

\vspace*{0.3cm}

\noindent\textbf{Text S2. Wind data processing}

The sand fluxes are derived from the wind data of the KHMN weather station at the Holloman air base ($32^{\circ}51'$N, $106^{\circ}06'$W) corresponding to the topography measurements, between 2007 and 2010. The method of analysis used here is based on that described in the supplementary material of \citeA{Cour14}. The data are hourly measurements of the average wind velocity $u_{t}$ and direction $\theta_{t}$ at $10$~m height, with velocity and angle bins of $0.5$~m/s and $10^{\circ}$, respectively. The sub-letter $t$ indicates here the time of the corresponding data.

The sediment flux depends on the shear velocity $u_{*}$, characteristic to the logarithmic velocity vertical profile of the wind $u(z) = u_*/\kappa \ln \left( z/z_{0}\right)$ inside the turbulent boundary layer. $\kappa=0.4$ is the von K\'arm\'an constant and $z_0$ is the aerodynamic roughness. At time $t$, this law of the wall is reciprocally used to estimate the shear velocity from the wind measurement as:
\begin{equation}
  u_{*, t} = \kappa\frac{u_t}{\ln\left( \frac{z}{z_{0}}\right)},
\end{equation}
where $z$ is set to the height at which the wind data have been recorded. Here we take $z_0 \simeq 1$~mm, associated with the thickness of the saltation layer.

Furthermore, the wind shear velocity exceeds a threshold $u_{\rm th}$ to maintain steady saltation. Its value is empirically given by
\begin{equation}
  u_{\rm th} = 0.1\sqrt{\frac{\rho_{\rm p}}{\rho_{\rm f}}gd},
\end{equation}
where $\rho_{\rm p}$ and $\rho_{\rm f}$ are the grain and air density, $d$ the grain diameter and $g$ the gravitational acceleration \cite{Iver99}. The saturated flux can then be computed at each time step using the transport law  \cite{Unga87}:
\begin{equation}
  \boldsymbol{q}_{{\rm sat}, t} = \begin{cases}
  \Omega \left(u_{*, t}^{2} - u_{\rm th}^{2}\right)\boldsymbol{e}_{\theta_{t}} & \textrm{if} \quad u_{*, t} > u_{\rm th}, \\
  0 & \textrm{else},\end{cases}
\end{equation}
where $\boldsymbol{e}_{\theta_{t}} = \left(\cos\theta_{t}, \sin\theta_{t} \right)$ is a unit vector, and $\Omega = 25\left(\rho_{\rm f}/\rho_{\rm p}\right)\sqrt{d/g}$ is a dimensional constant calibrated by measurements and numerical simulations (see review of \citeA{Dura11}). From this time series, one can define the resultant sand flux. Likewise, we also compute the characteristic sand flux $Q$ as:
\begin{equation}
  \boldsymbol{Q}_{t} = \begin{cases}
  \Omega \, u_{*, t}^{2}\boldsymbol{e}_{\theta_{t}} & \textrm{if} \quad u_{*, t} > u_{\rm th}, \\
  0 & \textrm{else}.\end{cases}
\end{equation}
Finally, as the sand flux direction is mostly unimodal, dune growth and migration mainly result from the component of the fluxes aligned with the resultant sand flux direction $\beta$, perpendicular to the dune crests. The relevant characteristic sand flux for the study of the spatial development of the dune instability then become:
\begin{equation}
  Q_{t} = \begin{cases}
  \Omega \, u_{*, t}^{2}\cos\left(\theta_{t} - \beta \right) & \textrm{if} \quad u_{*, t} > u_{\rm th}, \\
  0 & \textrm{else}.\end{cases}
  \label{Qt}
\end{equation}
Note that $Q_{t}$ can have positive or negative values depending on the orientation of the sand flux at time with respect to the resultant (average) sand flux direction. Quantification of the impact of the variations in wind direction can then be done by looking at the ratio between scalar and vector average of the characteristic sand flux. Using the wind data we obtain
\begin{align}
  \langle \vert Q_{t} \vert \rangle & = 38 \pm 5 \textrm{~m}^{2}~\textrm{yr}^{-1}, \\
  \langle Q_{t} \rangle & = 23 \pm 3 \textrm{~m}^{2}~\textrm{yr}^{-1},
\end{align}
resulting in
\begin{equation}
  \frac{\langle Q_{t} \rangle}{\langle \vert Q_{t} \vert \rangle} \simeq 0.6.
\end{equation}

\vspace*{0.3cm}

\noindent\textbf{Text S3. Linear stability analysis}

We derive here the linear stability analysis of a flat sedimentary bed sheared by a constant fluid flow, detailing the differences between temporal and spatial analyses. The calculations closely follow the framework presented in the review of \citeA{Char13}, leading to the characteristics of the most unstable mode.

As all unstable modes are found to be propagating downwind, the instability is said to be \emph{convective}: it grows while propagating. Then, only for an artificial situation with periodic boundary conditions, for example in numerical simulations, the instability can simultaneous grow in time, everywhere. Likewise, if one could flatten the sand bed at a kilometer scale, one would see the spatially-homogeneous emergence of the instability in time. Nevertheless, the dunes generated from noise at the upstream boundary condition would eventually invade, by propagation, a larger and larger domain.

\textit{General framework and notations} ---
The evolution of the bed elevation profile $h(x,t)$ is governed by the mass conservation (Exner) equation
\begin{equation}
\partial_t h + \partial_x q = 0,
\end{equation}
where $q$ is the sediment flux. In the steady homogenous case, the sediment flux takes its saturated value, which is a linear function of the basal shear stress $\tau_b \equiv \rho_f u_*^2$
\begin{equation}
q_{\rm sat} = \frac{\Omega}{\rho_f} \left( \tau_b - \tau_{\rm th} \right),
\end{equation}
where $ \tau_{\rm th} \equiv \rho_f u_{\rm th}^2$ is the threshold value of the stress, below which transport vanishes. In non-homogeneous situations, the flux lags behind its saturated value, following the relaxation equation
\begin{equation}
L_{\rm sat} \partial_x q = q_{\rm sat} - q,
\end{equation}
where $L_{\rm sat} $ is the saturation length.

In the context of the linear stability analysis, we consider a bed perturbation of the form
\begin{equation}
h(x,t) = H e^{i\left(k x - \omega t\right)},
\end{equation}
where where $k$ is the wavenumber and $\omega$ the associated pulsation. $H$ is the amplitude of the perturbation, and we always assume to work in the limit $kH \ll 1$. On such a sinusoidal bed, the associated basal shear perturbation writes in Fourier notation:
\begin{equation}
{\hat{\tau}}_b = \rho_f u_*^2 \left( \mathcal{A} + i \mathcal{B} \right) kH.
\end{equation}
$\mathcal{A}$ and $\mathcal{B}$ are hydrodynamic coefficients representing the in-phase and in-quadrature part of the shear stress with respect to the topography. Grouped together, the above equations lead to the following dispersion relation:
\begin{equation}
\frac{L_{\rm sat}^{2}}{Q}\omega = \displaystyle\frac{\left(L_{\rm sat }k\right)^{2}}{1 + i L_{\rm sat}k} \left(\mathcal{A} + i \mathcal{B}\right),
  \label{dispersion_relation}
\end{equation}
where $Q = \Omega u_*^2$ is the characteristic sand flux. If one accounts for the dependence of the threshold on the bed slope $\partial_x h$, one should simply replace the coefficient $\mathcal{B}$ by:
\begin{equation}
{\mathcal{B}}_\mu = \mathcal{B} - \displaystyle\frac{1}{\mu}\left(\frac{u_{\rm th}}{u_{*}}\right)^{2},
\end{equation}
where $\mu$ is the avalanche slope.

Generally speaking, the wavenumber and pulsation are complex numbers:
\begin{align}
k &= k_{r} + i k_{i}, \\
\omega &= \omega_{r} + i \omega_{i},
\end{align}
whose real parts ($k_{r}, \, \omega_{r}$) represent spatial and temporal oscillations, and whose imaginary parts ($k_{i}, \, \omega_{i}$) represent growth or decay of the amplitude of the perturbation. For sake of clarity, the pulsation and wavenumber will thereby only refer to their real parts, while one defines the temporal and spatial growth rates from their imaginary parts. From now on, the notations will then be the following:
 \begin{align}
    k_{r} &\equiv  k, \\
    \omega_{r} &\equiv \omega, \\
    \omega_{i}  &\equiv \sigma,\\
    -k_{i}  &\equiv 1/\Lambda,
\end{align}
where $\Lambda$ is the characteristic growth length. We also define the propagation velocity of the pattern as $c = \omega/k$.

\begin{figure}
  \centering
  \noindent\includegraphics[scale = 1, draft = false]{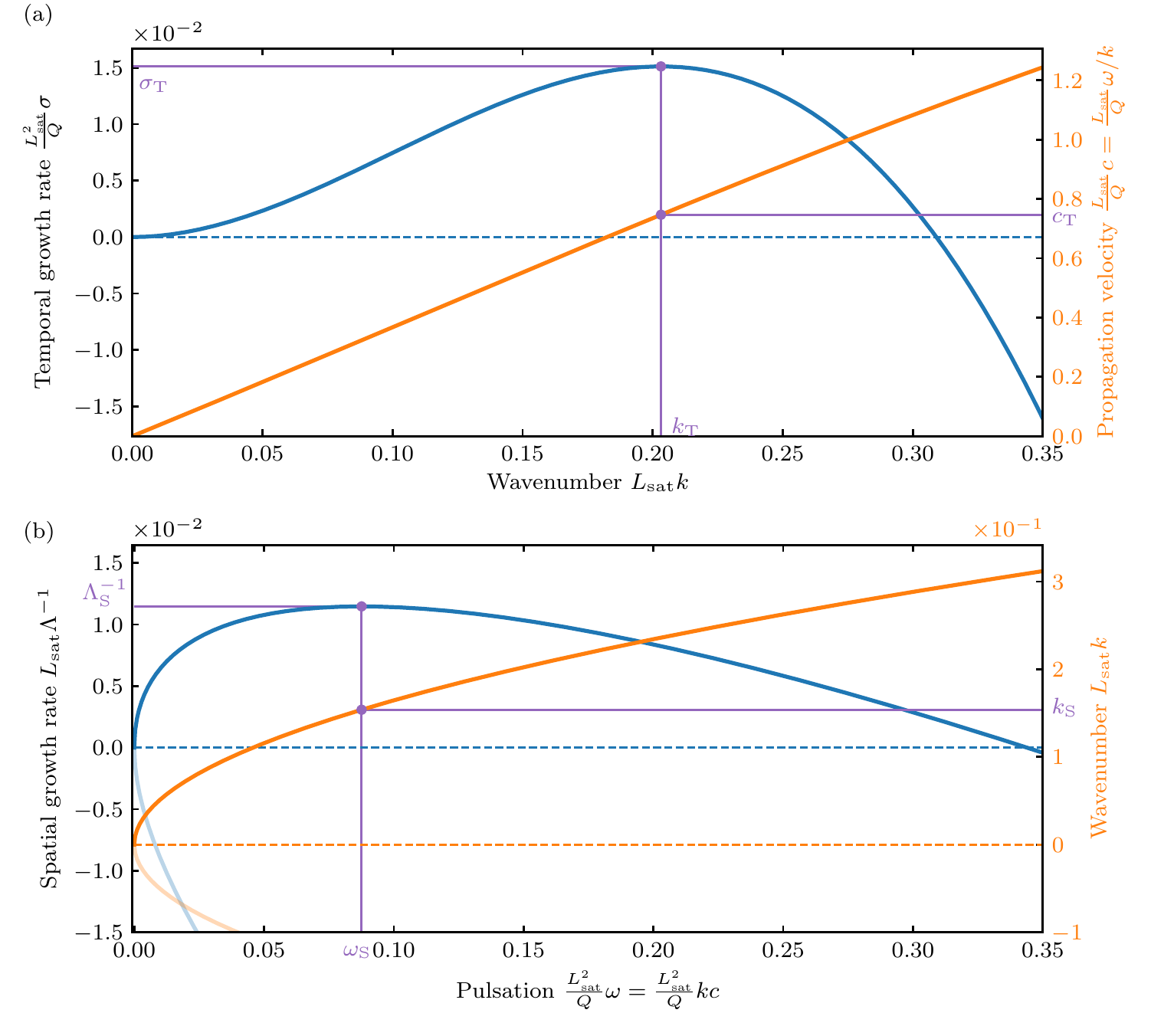}
  \caption{Dispersion relations corresponding to (a) the temporal linear stability analysis and (b) the spatial linear stability. Purple lines and dots indicates the most unstable mode. In (b), opaque and transparent curves correspond to the positive and negative branches $k_{+}$ and $k_{-}$, relating to upstream and downstream propagating waves, respectively. Here, $\mathcal{A} = 3.6$, $\mathcal{B} = 1.9$, $Q = 38$~m$^{2}$~yr$^{-1}$, $L_{\rm sat} = 2.8$~m, $\mu = 0.8$ and $u_{*}/u_{\rm th} = 1.26$.}
  \label{Figure_LSA}
\end{figure}

\textit{Temporal linear stability analysis} ---
We first analyse the temporal stability of system. We consider a spatially infinite system, where we disturb the bed at a time $t=0$ by a spatial oscillation of given wavelength $\lambda = 2\pi/k$. The system remains spatially homogeneous, which implies $1/\Lambda = 0$: there is neither spatial growth nor decay. We therefore look for the temporal response of the system to this perturbation, to compute its growth rate. The real and imaginary parts of the dispersion relation \eqref{dispersion_relation} can be expressed as:
\begin{align}
\frac{L_{\rm sat}}{Q} c(k)  &= \displaystyle\frac{ L_{\rm sat}k}{1 + \left(L_{\rm sat}k\right)^{2}} \left(\mathcal{A} + {\mathcal{B}}_\mu L_{\rm sat}k\right), \\
  \frac{L_{\rm sat}^{2}}{Q}\sigma(k)  &= \displaystyle\frac{ \left(L_{\rm sat}k\right)^{2}}{1 + \left(L_{\rm sat}k\right)^{2}} \left({\mathcal{B}}_\mu - \mathcal{A}L_{\rm sat}k\right).
\end{align}
Now we find the most unstable wavenumber $k_{\rm T}$ (Figure~\ref{Figure_LSA}(a)) corresponding to the wavelength $\lambda_{\rm T} = 2 \pi/k_{\rm T}$ at the maximum $\sigma$, assuming that $\mathcal{A}$ and $\mathcal{B}$ are constant (they are in fact weak (logarithmic) functions of $k$ in the rough turbulent regime of interest here \cite{Four10}). This most unstable wavelength writes in the limit $L_{\rm sat}k \ll 1$:
\begin{equation}
  \lambda_{\rm T} \sim \displaystyle\frac{3\pi\mathcal{A}}{{\mathcal{B}}_\mu} L_{\rm sat}.
  \label{lambda}
\end{equation}
The corresponding growth rate and propagation velocity are:
\begin{align}
  \sigma_{\rm T} &= \sigma(k_{\rm T}) \sim \displaystyle\frac{Q}{L_{\rm sat}^{2}} \frac{\mathcal{A}}{2} \left(\displaystyle\frac{2 \pi }{\lambda_{\rm T}}L_{\rm sat}\right)^{3} = \displaystyle \frac{Q}{L_{\rm sat}^{2}} \frac{4 {\mathcal{B}}_\mu^3}{27 \mathcal{A}^2}, \label{sigmaT} \\
  c_{\rm T} &= c\left(k_{\rm T}\right) \sim \displaystyle\frac{Q}{L_{\rm sat}} \mathcal{A} \left( \displaystyle\frac{2 \pi}{\lambda_{\rm T}}L_{\rm sat} \right) = \displaystyle\frac{Q}{L_{\rm sat}} \frac{2 {\mathcal{B}}_\mu}{3}. \label{cT}
\end{align}
The Lagrangian growth length can be defined as $c_{\rm T} / \sigma_{\rm T}$.

\textit{Spatial linear stability analysis} ---
Likewise, one can investigate the spatial stability of system. We perturb in time a specific point $x = 0$ of the sediment bed at a given pulsation $\omega$. As the system is temporally homogeneous at the beginning, so is its response, which implies $\sigma = 0$ (no temporal growth or decay). We therefore look for the spatial response of the system to this perturbation, to compute its growth length. The dispersion relation \eqref{dispersion_relation} can be rewritten as:
\begin{equation}
\left( \mathcal{A} + i{\mathcal{B}}_\mu \right)L_{\rm sat}^{2}\left(k - i\Lambda^{-1}\right)^{2} -i\frac{L_{\rm sat}^{2}}{Q}\omega L_{\rm sat}\left(k - i\Lambda^{-1}\right) - \frac{L_{\rm sat}^{2}}{Q}\omega = 0,
\end{equation}
which solves into:
\begin{equation}
  L_{\rm sat}\left(k - i\Lambda^{-1}\right) = \frac{1}{2\left(\mathcal{A} + i{\mathcal{B}}_\mu\right)}\left[i\frac{L_{\rm sat}^{2}}{Q}\omega \pm \sqrt{\frac{L_{\rm sat}^{2}}{Q}\omega \left[4\left(\mathcal{A} + i{\mathcal{B}}_\mu \right) - \frac{L_{\rm sat}^{2}}{Q} \omega\right]} \right].
  \label{dispersion_relation_spatial}
\end{equation}
The two roots $k_{+}$ and $k_{-}$ relate to waves propagating downstream and upstream, respectively. Splitting the above expression into real and imaginary part, we obtain:
\begin{align}
  L_{\rm sat} k(\omega) &= \displaystyle\frac{1}{2\left(\mathcal{A}^{2} + {\mathcal{B}}_\mu^{2}\right)}\left( {\mathcal{B}}_\mu \frac{L_{\rm sat}^{2}}{Q}\omega \pm  \mathcal{F}\left(\omega\right) \left[ \mathcal{A}\mathcal{G}\left(\omega\right) + \mathcal{B}_\mu\mathcal{H}\left(\omega\right) \right]\right), \\
  L_{\rm sat}\Lambda^{-1}(\omega) &= \displaystyle\frac{1}{2\left(\mathcal{A}^{2} + {\mathcal{B}}_\mu^{2}\right)}\left( -\mathcal{A} \displaystyle\frac{L_{\rm sat}^{2}}{Q}\omega \pm  \mathcal{F}\left(\omega\right) \left[ {\mathcal{B}}_\mu \mathcal{G}\left(\omega\right) - \mathcal{A}\mathcal{H}\left(\omega\right)\right]\right),
\label{dispersion_relation_spatial_split}
\end{align}
where
\begin{align}
\mathcal{F}\left(\omega\right) &= \sqrt{\displaystyle\frac{1}{2}\frac{L_{\rm sat}^{2}}{Q}\omega\left(4\mathcal{A} - \frac{L_{\rm sat}^{2}}{Q}\omega\right)}, \\
\mathcal{G}\left(\omega\right) &= \sqrt{1 + \sqrt{ 1 + \left(\displaystyle\frac{4{\mathcal{B}}_\mu}{4 \mathcal{A} - \displaystyle\frac{L_{\rm sat}^{2}}{Q}\omega}\right)^{2}}}, \\
\mathcal{H}\left(\omega\right) &= \sqrt{\sqrt{ 1 + \left(\displaystyle\frac{4{\mathcal{B}}_\mu}{4 \mathcal{A} - \displaystyle\frac{L_{\rm sat}^{2}}{Q}\omega}\right)^{2}} - 1}.
\end{align}
The spatial growth rate $1/\Lambda$ is plotted for both branches in Figure~\ref{Figure_LSA}b as a function of the pulsation $\omega$. Only the positive branch $k_{+}$ corresponding to waves propagating downstream exhibits positive values, and can grow. It also shows a single maxima, corresponding to the most unstable pulsation $\omega_{\rm S}$. The corresponding wavelength, propagation velocity and growth length are:
\begin{align}
  \lambda_{\rm S} & = \displaystyle\frac{2 \pi}{k\left(\omega_{\rm S}\right)}, \label{lambdaS} \\
  c_{\rm S} & = \displaystyle\frac{\omega_{\rm S}}{k\left(\omega_{\rm S}\right)}, \label{cS}\\
  \Lambda_{\rm S} & = \Lambda\left(\omega_{\rm S}\right). \label{LambdaS}
\end{align}
Finally, the Lagrangian growth rate is defined as:
\begin{equation}
\varsigma_{\rm S} \equiv c_{\rm S}/\lambda_{\rm S}.
\label{varsigmaS}
\end{equation}

\vspace*{0.3cm}
\noindent\textbf{Text S4. Time variations of the characteristic sand flux}

\begin{figure}
  \centering
  \noindent\includegraphics[scale = 1, draft = false]{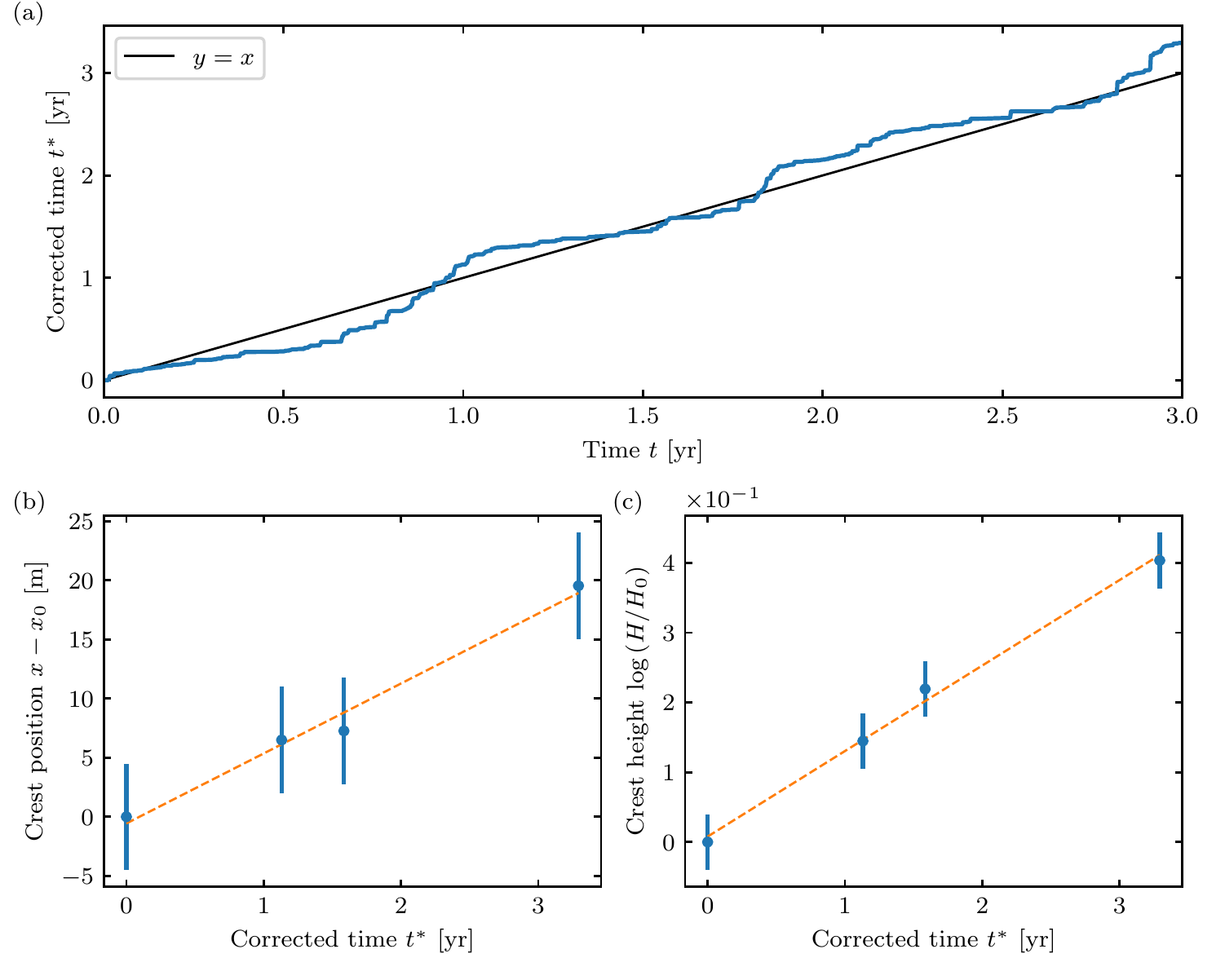}
  \caption{(a) Corrected time as a function of real time. (b) Position of a bump crest with respect to time. The dashed line is a linear fit giving $c = 5.9~\textrm{m}~\textrm{yr}^{-1}$. (c) Height log-ratio of a bump crest with respect to time. The dashed line is a linear fit giving $\varsigma = 0.012$~yr$^{-1}$. Errorbars come from the uncertainty due to the procedure for peak detection.}
  \label{Figure_sand_flux}
\end{figure}

The Lagrangian growth rate $\varsigma$ and the propagation velocity $c$ are obtained by fitting exponential and linear functions to the temporal variation of the dune height and position, respectively. However, both are also time-dependent through the characteristic sand flux $Q$ and the slope effect $(1/\mu)(u_{\rm th}/u_{*})^{2}$. In the fitting procedure to extract $\varsigma$ and $c$, we neglect this second contribution, but take into account the temporal variations of $Q$. Following \eqref{sigmaT} and \eqref{cT}, the Lagrangian growth rate and propagation velocity are proportional to the characteristic sand flux, such that we can write at each time step:
\begin{align}
  \varsigma_{t} = \displaystyle\frac{Q_{t}}{L_{\rm sat}^{2}} \bar{\varsigma}, \\
  c_{t} = \displaystyle\frac{Q_{t}}{L_{\rm sat}} \bar{c},
\end{align}
where $\bar{\varsigma}$ and $\bar{c}$ are non dimensional, assumed to be independent of time. Following the exponential growth and the linear propagation of a bump in the linear regime of the instability, its height $H$ and position $x$ at a time step $t$ can be expressed as:
\begin{align}
  H_{t} & = H_{0}\exp\left( \Sigma_{t} \varsigma_{t} \delta t \right), \\
  x_{t} & = x_{0} + \Sigma_{t} c_{t} \delta t,
\end{align}
which can be rewrittten as:
\begin{align}
  \ln\left(\frac{H_{t}}{H_{0}}\right) & = \displaystyle\frac{\langle Q_{t} \rangle}{L_{\rm sat}^{2}} \bar{\varsigma} \Sigma_{t} \displaystyle\frac{Q_{t}}{\langle Q_{t} \rangle} \delta t = \langle \varsigma \rangle t^{*}, \\
  x_{t} - x_{0} & =   \displaystyle\frac{\langle Q_{t} \rangle}{L_{\rm sat}} \bar{c} \Sigma_{t} \displaystyle\frac{Q_{t}}{\langle Q_{t} \rangle} \delta t = \langle c \rangle t^{*},
\end{align}
where $\langle \cdot \rangle$ denotes the time average. We then see that an effective time $t^{*}$ arises:
\begin{equation}
  t^{*} = \Sigma_{t} \displaystyle\frac{Q_{t}}{\langle Q_{t} \rangle} \delta t_{t}.
\end{equation}
It allows the extraction of the time average growth rate and propagation velocity (see Figure~\ref{Figure_sand_flux}), in order to be compared to the theoretical predictions, in which the average characteristic sand flux should be used.

\vspace*{0.3cm}

\noindent\textbf{Text S5. Distributions of the hydrodynamic coefficients and uncertainties}

\begin{figure}
  \centering
  \noindent\includegraphics[scale = 1, draft = false]{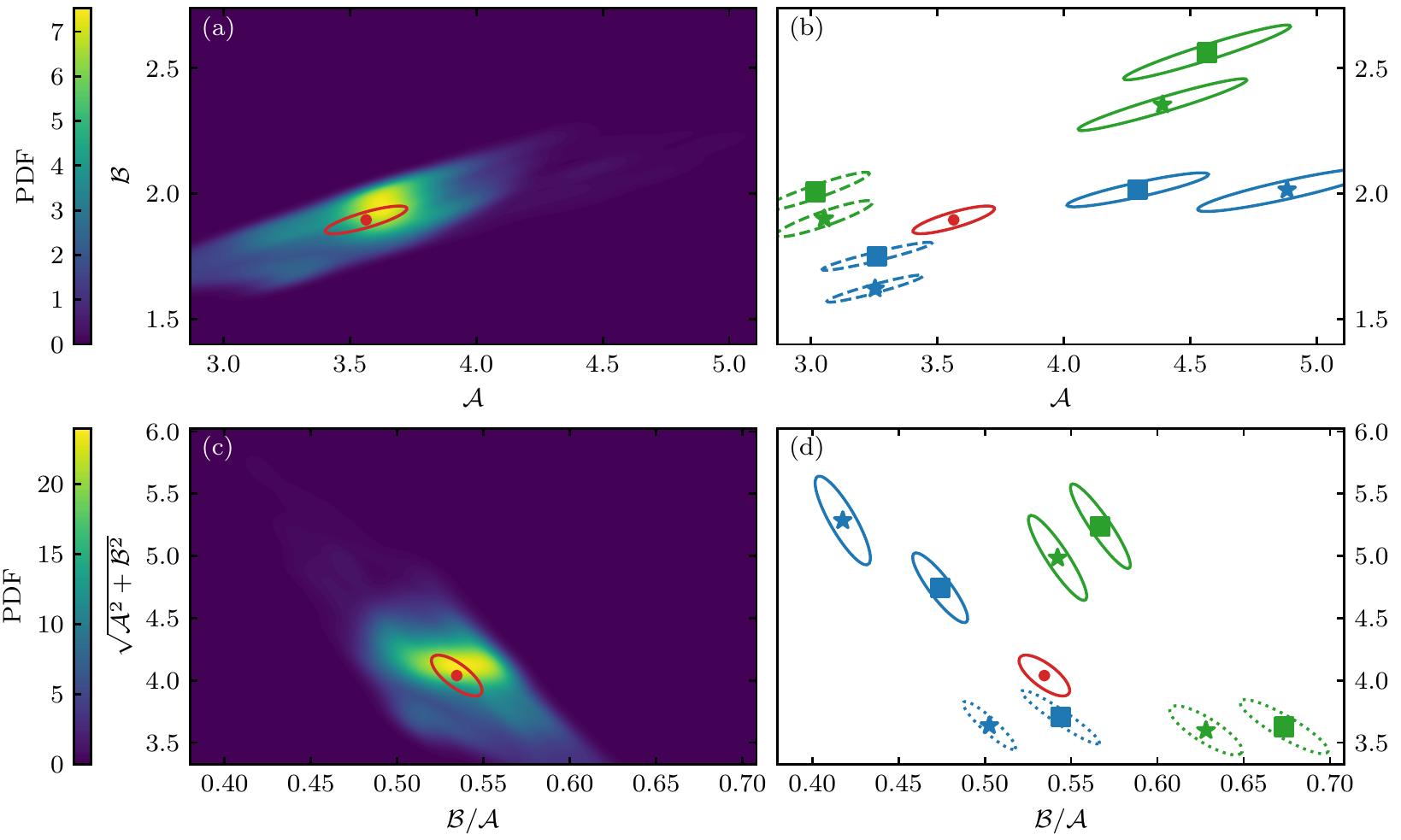}
  \caption{Distributions of the hydrodynamic parameters resulting from the inversion process. (a) Contour lines of the distribution when the parameters $Q$, $L_{\rm sat}$, $\mu$ and $u_{*}/u_{\rm th}$ are fixed to their mean value (Table~S1). The markers represent the mean value, and the ellipse the confidence interval at $95 \%$. (b) Variation of this ellipse when taking into account the uncertainty of the parameters given in Table~S1. Blue and green ellipses are for minimum and maximum plausible values of $L_{\rm sat}$, respectively. Plain and dashed ellipses are for minimum and maximum plausible values of flux $Q$. Squares and stars are the mean values of the distributions for minimum and maximum plausible values of the slope effect $(1/\mu)(u_{\rm th}/u_{*})^{2}$. (c) and (d) are the same than (a) and (b), but represented in the space spanned by the modulus $\sqrt{\mathcal{A}^2+\mathcal{B}^2}$ and the ratio $\mathcal{B}/\mathcal{A}$.}
  \label{Figure_inversion}
\end{figure}

In order to compare the predictions of the spatial linear stability analysis to the field data, we need to find the hydrodynamic coefficients $\mathcal{A}$ and $\mathcal{B}$. The field data are distributed around central values (Figure 3 of the main article), due to the intrinsic variability in dune fields. It is important to note that linear stability analysis predicts such a variability: the initial noise is selectively amplified in a range of wavelengths around the maximum growth rate. Here, we perform an inversion analysis and we determine the most probable values of $\mathcal{A}$ and $\mathcal{B}$, given the observations, using a simple method. We consider values of wavelength, growth length, Lagrangian temporal growth rate and propagation velocity inside the following intervals, which are representative of the peaks of the distributions:
\begin{align}
  \lambda &\in \left[100, 130\right]\textrm{~m}, \\
  \Lambda &\in \left[90, 250\right]\textrm{~m}, \\
  \varsigma &\in \left[0.0005, 1\right]\textrm{~yr}^{-1}, \\
  c &\in \left[3.5, 6.1\right]~\textrm{m}~\textrm{yr}^{-1}.
\end{align}
From a quadruple $\left\{ \lambda, \Lambda, \varsigma, c \right\}$, we compute the best tuple  $\left\{\mathcal{A}, \mathcal{B} \right\}$, taking the optimum prediction of equations (\ref{lambdaS}-\ref{varsigmaS}). This method results in a distribution for the hydrodynamic coefficients $\mathcal{A}$ and $\mathcal{B}$, determined from the statistical variability of the dune field. It exhibits clear dominant values, slightly different from the average ones due to the distribution shape. By taking the number of independent measurements into account, the error bars on $\mathcal{A}$ and $\mathcal{B}$ can be graphically represented by the ellipse of figure~\ref{Figure_inversion}(a). The tilt of the ellipse clearly shows that the hydrodynamic coefficients covary, which indicates a better precision on the ratio $\mathcal{B}/\mathcal{A}$ than on the individual values of $\mathcal{A}$ and $\mathcal{B}$.

\begin{figure}
  \centering
  \noindent\includegraphics[scale = 1, draft = false]{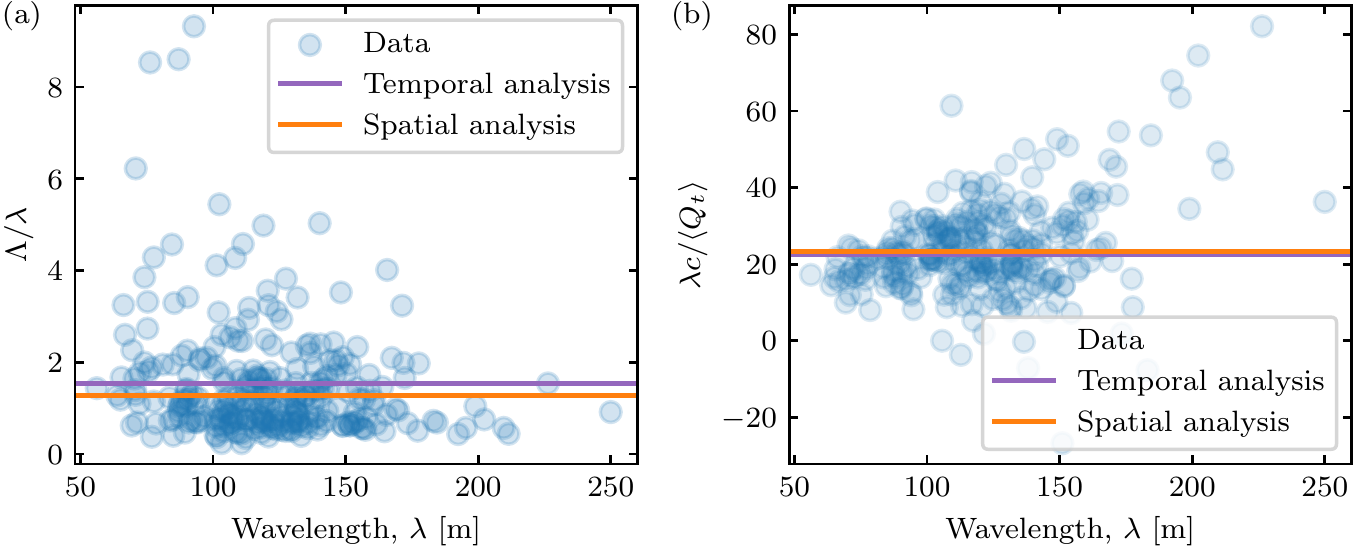}
  \caption{Pertinent non-dimensional numbers in the inversion process. (a) Ratio between the wavelength and the growth length. Following the temporal analysis, $\Lambda/\lambda \propto \mathcal{A}/\mathcal{B}_{\mu}$. (b) Ratio between the wavelength and the propagation length $\langle Q_{t}\rangle /c$. Following the temporal analysis, $\lambda c /\langle Q_{t}\rangle \ \propto \mathcal{A}$. The plain lines represent the predictions of the linear stability analysis for the parameter values given in Table~S1, with  $\mathcal{A} = 3.6$ and $\mathcal{B} = 1.9$.}
  \label{Figure_ratios}
\end{figure}

Average and most probable values also depend on the fixed parameters $Q$, $L_{\rm sat}$, $\mu$ and $u_{*}/u_{\rm th}$, whose uncertainties are given in Table~S1. Taking those uncertainties into account results in multiple different distributions (see for example Figure~\ref{Figure_inversion}b). Note that all the corresponding ellipses have roughly the same sizes, fixed by the dispersion of the measured quantities. The ratio $\mathcal{B}_\mu/\mathcal{A}$ is robustly determined as it is in principle related to both $\lambda/L_{\rm sat}$ and $\Lambda/L_{\rm sat}$. In particular, the theory predicts that the ratio
$\Lambda/\lambda \propto \mathcal{A}/\mathcal{B}_{\mu}$ should be constant, and independent of any other parameter.  Figure~\ref{Figure_ratios} shows that the field data nicely obey this property. The theory also predicts that the
ratio $\lambda c$ should be proportional to $\langle Q_{t}\rangle \mathcal{A}$. As a consequence, any uncertainty on $\langle Q_{t}\rangle$ results into an uncertainty on $\mathcal{A}$. Figure~\ref{Figure_ratios} shows that the ratio $\lambda c/\langle Q_{t}\rangle$ is, as predicted, independent of $\lambda$.

We therefore have two different sources of uncertainty; one resulting from the dispersion of the measured quantities, and one from the uncertainty on the fixed parameters. The value of $L_{\rm sat}$ is required to calculate $\lambda$ and $\Lambda$, and directly affects the estimate of $\mathcal{B}/\mathcal{A}$. Likewise, the value of $Q$ is required to calculate $c$, and mostly affects $\mathcal{A}$. Finally, the amplitude of the slope effect, $(1/\mu)(u_{\rm th}/u_{*})^{2}$ directly affects $\mathcal{B}$, as $\mathcal{B}_{\mu}$ remains the same.

The resulting uncertainty on hydrodynamic coefficients results from all contributions, leading to the final result: $\mathcal{A} = 3.6 \pm 0.6$ and $\mathcal{B} = 1.9 \pm 0.3$.

\bibliography{biblio.bib}

\end{article}

\clearpage

\begin{table}
\centering
\caption{Table of the parameters used for the linear stability analysis at the White Sands dune field. The star denotes direct measurements, while the others quantities are derived from measurements using calibrated laws. The air density is the value at $20^{\circ}$C, and the corresponding error includes corrections for elevation as well as for temporal variations in humidity and temperature.}
\vspace{.5cm}
\begin{tabular}{l c c c}
\hline
 Parameter  & Notation & Value & Unit  \\
\hline
Grain diameter* & $d$ & $670 \pm 120 $ & $\mu\textrm{m}$ \\
Grain bulk density* & $\rho_{\rm p}$ &  $2300 \pm 100 $ & $\textrm{kg}/\textrm{m}^{3}$ \\
Avalanche slope* & $\mu$ & $0.8 \pm 0.05$ &\\
Air density & $\rho_{\rm f}$ & $ 1.2 \pm 0.1$ & $\textrm{kg}/\textrm{m}^{3}$ \\
\hline
Saturation length & $L_{\rm sat}$ & $2.8 \pm 0.5$ & m \\
Shear velocity ratio & $u_{*}/u_{\rm th}$ & $1.26 \pm 0.05$ & \\
Characteristic flux & $Q$ & $38 \pm 5$ & $\textrm{m}^{2}/\textrm{yr}$ \\
\end{tabular}
\end{table}
\vspace{1.5cm}

\begin{table}
  \caption{Measured and predicted values of the characteristics of the emerging dune pattern. The peak value corresponds to the maximum of the smoothed distribution. The peak width is estimated at $3/4$ of its height.}
  \vspace{.5cm}

  \begin{tabular}{l c c c c}
     & Wavelength $\lambda$ & Growth length $\Lambda$ & Lagrangian growth rate $\varsigma$ & Propagation velocity $c$  \\
         & (m) & (m) & (yr$^{-1}$) & ($\textrm{m}~\textrm{yr}^{-1}$)  \\
    \hline
    \multicolumn{5}{c}{\textbf{Data from global method}} \\
    \hline
    Peak value & 122 & 112 & 0.013 & 4.4\\
    Peak width & 19 & 70 & 0.036 & 1.5 \\
    \hline
    \multicolumn{5}{c}{\textbf{Data from `peak-to-peak' method}} \\
    \hline
    Peak value & 116 & 86 & 0.015   &  4.6  \\
    Peak width &  80 & 125& 0.06 & 3.2  \\
    \hline
    \multicolumn{5}{c}{\textbf{Predictions under a unidirectional wind (red line in Figure~3)}} \\
    \hline
    Mean & 115 & 245 & 0.032 & 7.6 \\
    Minimum & 105 & 204 & 0.018 & 5.9 \\
    Maximum & 134 & 330 &0.049 & 10.1 \\
    \hline
    \multicolumn{5}{c}{\textbf{Predictions with correction due to reversing winds  (orange line in Figure~3)}} \\
    \hline
    Mean & 115 & 147 & 0.032 & 4.5 \\
    Minimum & 105 & 122 & 0.018 & 3.5 \\
    Maximum & 134 & 197 & 0.049 & 6
  \end{tabular}
\end{table}